\documentclass[a4paper]{article}

\usepackage{INTERSPEECH2021}
\usepackage{hyperref}       % hyperlinks
\usepackage[dvipsnames]{xcolor}

\title{Optimal Transport-based Adaptation in Dysarthric Speech Tasks}
\name{Rosanna Turrisi$^{1,2,3}$, Leonardo Badino$^4$}
\address{
  $^1$Italian Institute of Technology,
  $^2$University of Ferrara,
  $^3$DIBRIS, University of Genoa,
  $^4$ PerVoice}
\email{trrrnn@unife.it, leonardo.badino@pervoice.it}

\begin{document}

\maketitle
\begin{abstract}
In many real-world applications, the mismatch between distributions of training data (\textit{source}) and test data (\textit{target}) significantly degrades the performance of machine learning algorithms. In speech data, causes of this mismatch include different acoustic environments or speaker characteristics. In this paper,
we address this issue in the challenging context of dysarthric speech, by multi-source domain/speaker adaptation (MSDA/MSSA). Specifically, we propose the use of an optimal-transport based approach, called MSDA via Weighted Joint Optimal Transport (MSDA-WDJOT). We confront the mismatch problem in dysarthria detection for which the proposed approach outperforms both the Baseline and the state-of-the-art MSDA models, improving the detection accuracy of 0.9 $\%$ over the best competitor method. We then employ MSDA-WJDOT for dysarthric speaker adaptation in command speech recognition. This provides a Command Error Rate relative reduction of 16$\%$ and 7$\%$ over the baseline and the best competitor model, respectively. Interestingly, MSDA-WJDOT provides a similarity score between the source and the target, i.e. between speakers in this case. We leverage this similarity measure to define a Dysarthric and Healthy score of the target speaker and diagnose the dysarthria with an accuracy of 95$\%$. 

In many real-world applications, the mismatch between   training data (\textit{source}) distribution and test data (\textit{target}) distribution significantly degrades the performance of machine learning algorithms.
%[Puoi anche togliere la frase seguente] In automatic speech recognition, causes of this mismatch include different acoustic environments or speaker characteristics. 
In this paper,
we address this problem in the context of assessment and recognition of dysarthric speech, in a multi-source domain/speaker adaptation (MSDA/MSSA) setting. %non so se setting e' la parola piu' appropriata qui, forse vanno scritte 2 parole in piu'.
Specifically, we propose the use of an optimal-transport based approach, called MSDA via Weighted Joint Optimal Transport (MSDA-WDJOT). We confront the mismatch problem in dysarthria detection for which the proposed approach outperforms both the [baseline non va con b maiuscola e devi spiegare in 2 parole cosa e' la baseline] Baseline and the state-of-the-art MSDA models, increasing detection accuracy by 0.9 $\%$ over the best competitor method. We then employ MSDA-WJDOT for dysarthric speaker adaptation in a command speech recognition task. Our approach reduces the command error rate by 16$\%$ and 7$\%$ over the baseline and the best competitor model, respectively. Interestingly, MSDA-WJDOT provides a similarity score between the source and the target, i.e. between speakers in this case. We leverage this similarity measure to define a Dysarthric and Healthy score of the target speaker and diagnose the dysarthria with an accuracy of 95$\%$. 

\end{abstract}

\section{Introduction}

Many machine learning algorithms assume that the test and training datasets are
sampled from the same distribution. However, in many real-world applications, new
data can exhibit a distribution change (\textit{domain shift}) that degrades the algorithm
performance. In speech tasks, such as speech recognition, this shift can be observed when the recording conditions are varying (e.g., different noisy/clean environments). This problem can be solved through Domain Adaptation (DA) \cite{Jiang2007, Wouter2019}, that is a particular case of transfer learning \cite{Pan2009}. DA methods consist in leveraging a labelled dataset, called \textit{source} domain, to build a model that well performs on the dataset of interest, called \textit{target} domain. The source and target domains are supposed to be similar and, typically, the labels for the target domain are not available (unsupervised DA). 

Beside the recording conditions mismatching, speech tasks suffer of differences between speakers (e.g., different accents, different speaking style). This mismatch is even more evident in speech-impaired people as the speech characteristics also depend on the type and the severity of the disorder. We refer to this problem as \textit{speaker adaptation}. \\

In this work, we focus on unsupervised domain adaptation and unsupervised speaker adaptation in pathological speech, when multiple sources are available. Therefore, in this framework, we assume to have access to multiple labelled training data sets (e.g., multiple speakers data) that are different, but related, to the dataset of interest (e.g., the speaker of interest) for which labels are not available. 
We consider as pathological speech the one recorded from people affected by dysarthria. This is a motor speech disorder consisting in the disruption of the normal control of the vocal tract musculature that is common in elderly people and in conditions such as Parkinson Disease, Amyotrophic Lateral Sclerosis and post-stroke motor impairments \cite{mcneil2009clinical, darley1969}.

We consider speech recorded from people with affected by dysarthria, that is a motor disorder consisting in the disruption of the normal control of the vocal tract musculature \cite{darley1969}. Dysarthria can express different speech characteristics based on the location of the neurological damage. However, it is possible to individuate common tendencies in dysarthric patients, such as mumbled speech, an acceleration or deceleration in the speaking rate, abnormal pitch and rhythm. Further, dysarthric individuals are more subject to a fatigue factor that affects the voice. We here investigate two case studies, that are domain adaptation in dysarthria detection and dysarthric speaker adaptation for command speech recognition.

Dysarthria detection is currently evaluated by neurology experts through the use of clinical assessment tools that attribute a score to the capacity of the subject to perform perceptual and/or acoustic tasks. %Even though the role of the experts still remains fundamental, the detection might be biased and subjective. For example, the final score of dysarthria severity is strictly linked to the choice of the clinical assessment tool. Further,
This procedure can be laborious and time consuming. Therefore, a rapid and objective dysarthria detection procedure could help the therapist in the diagnosis.

In the last years, the research community started to look at dysarthria detection by learning a mapping from the acoustic features to the text label \cite{williamson2015, an2015}. 
In \cite{tu2017}, the authors proposed an interpretable DNN model in which they added an intermediate layer that acts as a bottle-neck feature extractor providing nasality, vocal quality, articulatory precision and prosody features. They showed that this interpretable features are highly correlated with the evaluation of speech-language pathologists. In \cite{cernak2017} the authors propose a deep learning approach to compute phonological posteriors from the speech signal and model the voice quality spectrum in patient affected by Parkinson's Disease (PD).
In \cite{vasquez2017},  articulation impairments of PD patients are
analyzed by time–frequency representations (TFRs), used to feed a convolutional neural network (CNN) and discriminate between patients and healthy speakers.

However, none of the previous studies considered the mismatch between training
and testing data. Towards this direction, few studies \cite{wisler2014domain, Berisha} looked for an invariant-feature space to improve dysarhtria detection. Nonetheless, these only consider the mismatch case within the same dataset, in which recording conditions and environment characteristics can be considered invariant. Instead, we here tackle the more real and common scenario in which audio are recorded in different noisy conditions.
To the best of our knowledge, this is the first work in which unsupervised domain
adaptation is applied to dysarthria detection. While the domain
adaptation problem for dysarthric speech has not been addressed yet, the adaptation of
a model to a new unseen speech dataset with a different noise type is an old problem
that always attracted the attention of the research community \cite{van1989, cung1993, furui2008}. Due to the fact that both dysarthria detection and noise robustness are challenging tasks, they tend to be treated separately. However, it is worthwhile to consider them as a unique problem. Indeed, dysarthric speech corpora are usually collected at the
hospitals that are not equipped to record speech in absence of noise.

Speaker Adaptation (SA) techniques have been widely
investigated by the automatic speech recognition (ASR) community \cite{gemello2007, huang2015, zhao2016, sun2017, meng2018, meng2019}. However, 
conventional SA algorithms perform poorly in
dysarthric speech when they present a low intelligibility. As reported in \cite{kotler1997effects, manasse2000speech, raghavendra2001investigation},
even though commercial ASR systems can achieve up to 90$\%$ for some dysarthric
speakers with high intelligibility, the recognition performance still remains inadequate
with the decreasing of the speech intelligibility. These commercial systems usually
incorporate SA techniques to adapt the model to the voice of that speaker that require
some audio samples from the speaker of interest. It follows that traditional adaptation
techniques are not sufficient to deal with gross abnormalities \cite{green2003automatic}. Further, they
results to be inefficient even for speakers with mild to moderate dysarthria when the
vocabulary size is larger than 30 words \cite{goodenough1991towards}.

In the last decades, some attempts to move forward and solve the training-testing
mismatch for pathological speech have been done. \cite{mustafa2014} adapts the ASR system trained on large datasets to a dysarthric dataset. In \cite{yilmaz2018}, the authors leverage articulatory features, as well as the acoustic ones, achieving a 4-8$\%$ of World Error Rate (WER) relative reduction. However, these techniques require a large amount of data
for fine-tuning, that is usually not available. To overcome this problem, \cite{shor2019} proposes to fine-tune only a subset of the network layers to better adapt an ASR model to the dysarthric speech.  
\cite{morales2009}, rather than adapting the acoustic models, models the errors at phonetic level
made by the speaker and attempts to correct them by two possible strategies, that incorporate the language model and find the best estimate of the correct word sequence.\\
However, all the aforementioned studies are limited to the supervised framework,
whereas the unsupervised speaker adaptation is the most common real scenario.

We propose to solve both adaptation problems by adopting an Optimal Transport (OT) \cite{Monge, Kantorovich} based approach, named Multi-Source Domain Adaptation Weighted Joint Distribution Optimal Transport (MSDA-WJDOT), firstly introduced in \cite{turrisi2020multi}. This method looks for a convex combination of the joint source distributions with minimal Wasserstein distance to the proxy joint target distribution, in which the labels are replaced by the prediction of a classifier. The source weights and the target classifier are simultaneously learned.  The advantage of MSDA-WJDOT over the other MSDA approaches is that the source weights provide a similarity measure between each source and the target domain. This can be leveraged to select only the relevant sources and it also offers an interpretable model. For instance, in the case of speaker adaptation, these weights reflect the similarity between speakers.  

%This paper is organized as follow. Sec. \ref{sec:WJDOT} introduces MSDA-WJDOT method. Sec.  \ref{sec:Exp} describes the experimental settings, including the corpora description and the model details. Results are reported in Sec. \ref{sec:results}. Finally, a discussion is provided in Sec. \ref{sec:discussion}.

\section{MSDA-WJDOT} \label{sec:WJDOT}

In this section, we recall MSDA-WJDOT firstly proposed in \cite{turrisi2020multi}. This method approaches to the MSDA problem by estimating the similarity between the source and target domains, in the Wasserstein sense, and learning a target classifier using only the most similar source datasets.

Let us suppose to have $J$ sources $(X_{j}, Y_{j})$ with $N_{j}$ samples. MSDA-WJDOT assumes to have access to a differentiable embedding function $g$ from the input space to an embedding space $\mathcal{G}$. If the embedding is not available MSDA-WJDOT becomes a two-step procedure where  $g$ is learned in the first step. This can be done by  
\begin{equation}\label{eq:baseline}
 \operatorname*{min}_{g, f_{S}} \quad\sum _{j=1}^{J} \frac{1}{N_{j}} \sum _{i=1}^{N_{j}} \mathcal{L}(f_{S} \circ g(x_{j}^{i}), y_{j}^{i}),
\end{equation} 
where $f_{S}$ is a global classifier common to all sources. Alternatively, the authors in \cite{turrisi2020multi} suggests to estimate $g$ with the Multi-Task Learning (MTL) framework \cite{Caruana1997}, i.e. by
 \begin{equation}\label{eq:standardMTL}
 \operatorname*{min}_{g, \{f_{j}\}_{j=1}^{J}} \quad\sum _{j=1}^{J} \frac{1}{N_{j}} \sum _{i=1}^{N_{j}} \mathcal{L}(f_{j}\circ g(x_{j}^{i}), y_{j}^{i}),
\end{equation} 
where $f_{j}$ represents the classification function of the $j$-th source. Note that Eq. \ref{eq:baseline} and \ref{eq:standardMTL} provide two different ways to estimate $g$, while the source classifiers $f_S$ and $f_j$ are not further used in MSDA-WJDOT.

Once $g$ is given, we can define the source joint distributions $p_{S,j}$, for $1\leq j\leq J$, with support on the product space $\mathcal{G}\times\mathcal{Y}$ , where $\mathcal{Y}$ is the label space. We define a convex combination of the source distributions
$$    p_{S}^{\alpha} = \sum _{j=1}^{J} \alpha _{j}p_{S,j}$$
with $\pmb{\alpha}\in\Delta^{J}$. MSDA-WJDOT aims at finding a classification function $f$ that aligns the target distribution $p_T^{f}$ with a convex combination  $\sum _{j=1}^{J} \alpha _{j} p_{S,j}$ of the source distributions with convex weights $\pmb{\alpha}$ {on the simplex $\Delta^J$}. This can be expressed as 
\begin{equation}\label{eq:DA1}
 %\pmb{\alpha}^{*}, f^{*} = 
 \operatorname*{min}_{\pmb{\alpha}, f} \quad
W_D\left(\hat p_T^{f}, \sum _{j=1}^{J} \alpha _{j} \hat p_{S,j}\right),
\end{equation}

{where $W_{D}$ is the Wasserstein distance.}
A remarkable advantage of this approach is that the weights  $\pmb{\alpha}$ are learned simultaneously with the classifier $f$, which allows to distribute the mass based on the similarity between the source and the target domains, both in the feature and in the output spaces. Consequently, misleading and unrelated source domains are in practice not used and this allows to learn a more accurate classifier. 

In the following we refer to MSDA-WJDOT as  \texttt{MSDA-WJDOT} or \texttt{MSDA-WJDOT}$_{MTL}$ to specify if the embedding is learned by Eq. \ref{eq:baseline} or \ref{eq:standardMTL}, respectively.

% tab dei risultati - dysarthria detection
\begin{table*}[!ht]
\caption{Dysarthria detection accuracy on four target datasets: F16, Buccaneer2 (B2), Factory2 (F2), Destroyerengine (D). The mean and standard deviation of the accuracy are reported for the Baseline, IWERM, two extensions of JDOT and the proposed MSDA-WJDOT approach.}
\begin{center}
%\resizebox{\linewidth}{!}{% 
\begin{tabular}{c|c|c|c|c|c} 
\hline 
\textbf{Target domain} & \textbf{F16} & \textbf{B2} & \textbf{F2} & \textbf{D} & \textbf{Average}\\
\hline
\texttt{Baseline} & 93.59 $\pm$ 0.38 & 93.76 $\pm$ 0.22 & 93.23 $\pm$ 0.66 & 92.46 $\pm$ 0.82 & 93.26\\
\hline
\texttt{IWERM} \cite{Sugiyama2007} & 66.22 $\pm$ 0.01 & 66.38 $\pm$ 0.01 & 66.25 $\pm$ 0.05 & 66.30 $\pm$ 0.09 & 66.29 \\
\texttt{CJDOT}$_{MTL}$ \cite{Courty2017}& 95.35 $\pm$ 0.55 &97.39 $\pm$ 0.09 & 96.71 $\pm$ 0.07 & 92.98 $\pm$ 0.75& 95.61\\
\texttt{MJDOT}$_{MTL}$ \cite{Courty2017}& 95.81 $\pm$ 0.42 & 97.22 $\pm$ 0.09 & 96.53 $\pm$ 0.12 &93.12 $\pm$ 0.67& 95.67\\
\hline
\texttt{MSDA-WJDOT}$_{MTL}$ & \textbf{97.32} $\pm$ \textbf{0.36} & \textbf{97.82 $\pm$ 0.13} & \textbf{97.76}$\pm$ \textbf{0.10} & \textbf{95.42} $\pm$ \textbf{0.24} & \textbf{97.08} \\
\hline
\end{tabular}%}
\end{center}
\label{tab:detection}
\end{table*} 
%%%

\section{Experimental setup} \label{sec:Exp}
\subsection{Dysarthria detection}
\subsubsection{Dataset} TORGO is one of the most popular dysarthric speech corpora \cite{Rudzicz2012torgo}. It consists of aligned acoustic and articulatory recordings from 15 speakers. Seven of
these speakers are control speakers without any speech disorders, while the remaining eight speakers present different levels of dysarthria. We used the TORGO dataset to generate multiple-sources and target domain. In
particular, we generated 15 noisy datasets by combining the raw signals with different
types of noises from a noise dataset (available  \href{http://spib.linse.ufsc.br/noise.html}{\textit{here.}}). {Each noisy dataset has been synthesized by
PyDub python library \cite{pydub} summing the same type of noise signal to the raw data}. We then used the libROSA Python library \cite{librosa} to extract
13 MFCCs plus deltas and delta-deltas, computed every 10ms from 25ms Hamming
windows followed by a z-normalization per track.
We fixed the number of sources equal to 14 and we tested 4 noisy domains as target:
F16, Buccaneer2 (B2), Factory2 (F2), Destroyerengine (D). 

%\NB{It is also not clear to me whether there was any partitioning of speakers in the detection system. To properly evaluate the performance of a system like this it is essentially that distinct speaker groups for training and testing are distinct (otherwise the system could be learning speaker-specific patterns unrelated to the dysarthria). If this type of partitioning isn’t performed, it should be at the very least highlighted as a limitation of the experiments that are conducted.}

\subsubsection{MSDA-WJDOT model details} The feature extraction $g$ is performed by a Bidirectional Long Short-Term Memory (BLSTM) recurrent network with two hidden layers, each containing 50 memory blocks. We train the BLSTM as sequence-to-vector model, by taking the final hidden state at the last time step as a fixed-length utterance representation to generate the command probability given the whole sequence. The source and target classifier functions $\{f_{j}\}$,$f_{S}$, $f$ are learned as one feed-forward layer taking the last hidden state as input. The weights were initialized with Xavier initialization. Training is performed with Adam optimizer with 0.9 momentum and $\epsilon = e^{-8}$. The learning rate exponentially decays at every epoch. We grid-research the initial learning rate value and the decay rate. 

\subsection{Command speech recognition}
\subsubsection{Dataset} To investigate the Dysarthric Speaker Adaptation we employ the AllSpeak dataset \cite{dinardi}, that consists of speech recordings from 29 Italian native speakers. Seventeen of these (thirteen males, four females) are affected by Amyotrophic Lateral Sclerosis, while the remaining twelve (six males, six females) are healthy control speakers.  The dataset contains 25 commands in Italian, relative to basic needs such as ``I am thirsty". This dataset is very challenging due to the small amount of recordings. Indeed, only 2387 and 1857 examples have been recorded from control and dysarthric speakers, respectively. 

 We perform speaker adaptation of each dysarthric speaker by using all the remaining speakers as training dataset. The unlabelled target speaker data is split into adaptation set (80$\%$) and testing (20$\%$) set, which contains one example of each command. To train MSDA-WJDOT, we simultaneously employ the source training dataset and the adaptation target speaker dataset.

\subsubsection{MSDA-WJDOT model details}
The embedding function $g$ is represented by a BLSTM with five hidden layers, each containing 250 memory blocks. Finally, a softmax layer performs the classification task. Here, we do not consider the MTL variant as the dataset size is limited and learning a source-classifier $f_{j}$ with a very small amount of data may result hard. All weights are initialized with Xavier initialization. Training is performed with Adam optimizer with 0.9 momentum and $\epsilon = e^{-8}$. Learning rate is fixed to 0.001.

\section{Results} \label{sec:results}

\subsection{Competitor models}
We compare the proposed approach with well-established MSDA approaches: Importance Weighted Empirical Risk Minimization (\texttt{IWERM}) \cite{Sugiyama2007} and two extensions of JDOT \cite{Courty2017} to the MSDA case (see \cite{turrisi2020multi} for more details). To better measure the adaptation contribution, we also report the performance of a \texttt{Baseline} model in which the global classifier $f_{S}$, learned on the source domains, is used as target classifier. In addition, in Sec. \ref{sec:CSR}, we provide the performance of a supervised speaker adaptation (\texttt{SSA}) model, in which we add a feed-forward linear layer atop the input to the \texttt{Baseline} model and train it on the target dataset \cite{neto1995}. We consider this approach as the lower bound of the SA performance error.

\subsection{Dysarthria detection}
In all experiments, we observed that learning $g$ by the Multi-Task Learning approach always provides a better performance. Hence, for an easier reading of Table \ref{tab:detection}, we only report the performances in which the extractor $g$ is given by the MTL. 

For all the target domains, \texttt{IWERM} performs poorly, even underperforming  the \texttt{Baseline}. This is probably due to the difficulties in computing the probability density function of the acoustic input, that presents a high complexity. Indeed, to make the computation feasible, we firstly extracted a lower-dimensional vector from the audio signal by PCA. All the remaining MSDA methods outperform the Baseline. Among them our MSDA-WJDOT provides the best accuracy for all target domains. More precisely, MSDA-WJDOT provides a relative error reduction of 56.7$\%$ and 32.69$\%$ over the error of the \texttt{Baseline} and the best competitor model, respectively.

\subsection{Command speech recognition}\label{sec:CSR}
%\begin{table}[th]
%\caption{Command Error Rate (CER) for each dysarthric target speaker provided by the \texttt{Baseline}, \texttt{MSDA-WJDOT} and the competitor models.}
%\begin{center}
%\resizebox{\linewidth}{!}{% 
%\begin{tabular}{c|cccc} 
%\textbf{Speaker} &\texttt{Baseline} & \texttt{CJDOT} \cite{Courty2017} & \texttt{MJDOT} \cite{Courty2017} & \texttt{MSDA-WJDOT}\\
%\hline
%\textbf{M01} & 35.79 & 32.77 & \textbf{31.93} & \textbf{31.93} \\
%\textbf{M02}& \textbf{34.26} & 36.75 & 36.75 & 37.71\\ 
%\textbf{F01}& 63.16 & 52.63 & \textbf{49.12} &  \textbf{49.12} \\ 
%\textbf{F02}& 48.50 & \textbf{40.00} & \textbf{40.00} & \textbf{40.00} \\ 
%\textbf{M03} & 64.44 & 68.89 & 71.11 & \textbf{57.78}\\ 
%\textbf{M04} & \textbf{30.00} & 31.20 & 32.00 & 30.40 \\ 
%\textbf{M05} & 18.62 & 17.46 & 15.08 & \textbf{14.29} \\ 
%\textbf{F03} & 68.33 & \textbf{61.74} & 70.43 & 62.61 \\ 
%\textbf{M06} & 48.67 & 34.78 & \textbf{33.91} & 35.65\\ 
%\textbf{M07} & 11.00 & \textbf{7.20} & 8.00 &  8.80 \\ 
%\textbf{M08} & 39.50 & 36.00 & 41.60 & \textbf{33.60} \\ 
%\textbf{M09} & 24.79 & \textbf{16.81} & 18.49 & 19.33 \\ 
%\textbf{F04} & 48.07 & \textbf{38.60} & \textbf{38.60} & \textbf{38.60} \\ 
%\textbf{M10} & 18.00 & 12.80 & \textbf{12.00} & 12.80 \\ 
%\textbf{M11} & 56.50 & 47.20 & 48.80 & \textbf{45.60} \\ 
%\textbf{M12} & 7.50 & 5.60 & 7.20 & \textbf{4.80} \\ 
%\textbf{M13} & 30.91 & 45.45 & 43.64 & \textbf{21.82} \\ 
%\hline
%\textbf{A. CER} & 38.11& 34.46 & 34.37 & \textbf{32.05}\\
%\textbf{A. Rank} & 2.88 & 1.94 & 2.29  & \textbf{1.65} \\
%\end{tabular}}
%\end{center}
%\label{tab:SA}
%\end{table} 

\begin{table}[th]
\caption{Command Error Rate (CER) for each dysarthric target speaker provided by the \texttt{Baseline}, \texttt{SSA},  \texttt{MSDA-WJDOT} and the competitor models.}
\begin{center}
\resizebox{\linewidth}{!}{% 
\begin{tabular}{c||c|ccc|c} 
\textbf{Speaker} &\texttt{Baseline} & \texttt{CJDOT} \cite{Courty2017} & \texttt{MJDOT} \cite{Courty2017} & \texttt{MSDA-WJDOT} & \texttt{SSA}\\
\hline
\textbf{M01} & 35.79 & 32.77 & \textbf{31.93} & \textbf{31.93} & 25.00 \\
\textbf{M02}& 34.26 & \textbf{36.75} & \textbf{36.75} & 37.71 & 54.17\\ 
\textbf{F01}& 63.16 & 52.63 & \textbf{49.12} &  \textbf{49.12} & 60.00 \\ 
\textbf{F02}& 48.50 & \textbf{40.00} & \textbf{40.00} & \textbf{40.00} & 36.00 \\ 
\textbf{M03} & 64.44 & 68.89 & 71.11 & \textbf{57.78} & 55.56 \\ 
\textbf{M04} & 30.00 & 31.20 & 32.00 & \textbf{30.40} & 24.00\\ 
\textbf{M05} & 18.62 & 17.46 & 15.08 & \textbf{14.29} & 17.39 \\ 
\textbf{F03} & 68.33 & \textbf{61.74} & 70.43 & 62.61 & 56.52 \\ 
\textbf{M06} & 48.67 & 34.78 & \textbf{33.91} & 35.65 & 26.09 \\ 
\textbf{M07} & 11.00 & \textbf{7.20} & 8.00 &  8.80  & 20.00 \\ 
\textbf{M08} & 39.50 & 36.00 & 41.60 & \textbf{33.60} & 32.00 \\ 
\textbf{M09} & 24.79 & \textbf{16.81} & 18.49 & 19.33 & 13.04 \\ 
\textbf{F04} & 48.07 & \textbf{38.60} & \textbf{38.60} & \textbf{38.60} & 40.91 \\ 
\textbf{M10} & 18.00 & 12.80 & \textbf{12.00} & 12.80 & 16.00 \\ 
\textbf{M11} & 56.50 & 47.20 & 48.80 & \textbf{45.60} & 40.00\\ 
\textbf{M12} & 7.50 & 5.60 & 7.20 & \textbf{4.80} &4.00 \\ 
\textbf{M13} & 30.91 & 45.45 & 43.64 & \textbf{21.82} & 9.09\\ 
\hline
\textbf{A. CER} & 38.11& 34.46 & 34.37 & \textbf{32.05} & 31.16\\
\end{tabular}}
\end{center}
\label{tab:SA}
\end{table}

Table \ref{tab:SA} reports the results in terms of Command Error Rate (CER). A first remark is that, although the \texttt{Baseline} always achieved a CER between $15\%$ and $20\%$ on the validation set, it often had low accuracy on the target speaker. Once again, this emphasizes the difficulty of an ASR system to generalize to a new dysarthric speaker and the importance of the speaker adaptation in this context.  \\
The unsupervised speaker adaptation carried out by \texttt{MSDA-WJDOT} outperforms all the methods by providing the best Average CER. Indeed, it reduces the CER of $16\%$ and $7\%$ over the \texttt{Baseline} and the MSDA competitors, respectively. Surprisingly, MSDA-WDJOT achieves an Average CER similar to the \texttt{SSA} approach, in which the labels are used. 

%Further, we provide the Average Rank that is a performance measure suitable when several targets domain are tested. To every method, it assigns a rank (from 1 to 4, that is the number of considered methods) for each tested target based on the CER (e.g., 1 if the method has the lowest CER, 4 for the highest CER) and then it computes the average of the ranks. This measure is more robust to the variance and confirms that \texttt{MSDA-WJDOT} provides the best performance. \\

\begin{figure}[t]
\centering\includegraphics[width=.9\linewidth]{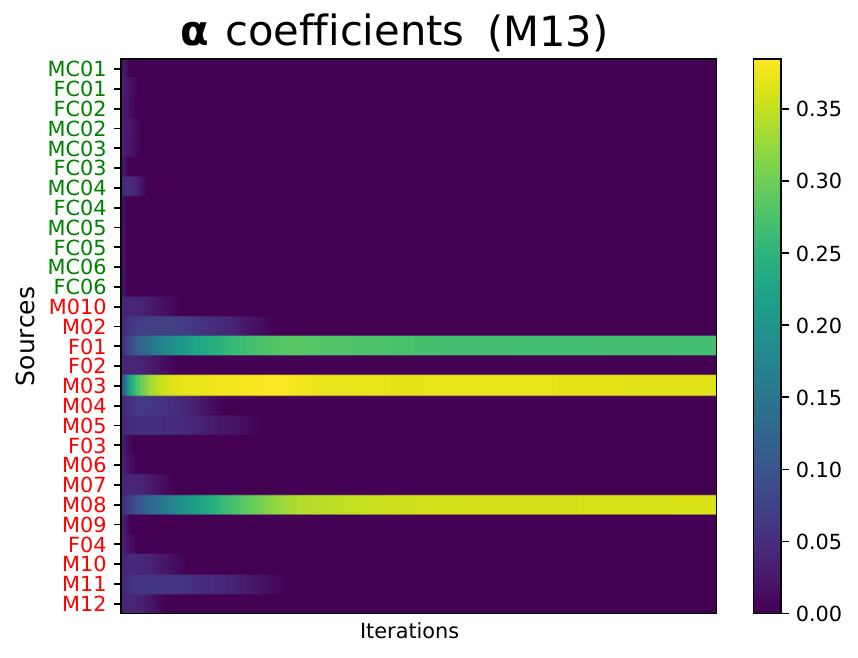}
\caption{$\pmb{\alpha}$ coefficients recovered during MSDA-WJDOT training for dysarthric target speaker M13. The $\alpha _{j}$ coefficients are close to zero for healthy speakers (in green), while the highest weights are attributed to dysarthric speakers (in red)}
\label{fig:alpha_M13}
\end{figure}

It is crucial to recall that \texttt{MSDA-WJDOT} provides a measure of similarity between the target and the sources and, hence, between speakers. We found that the recovered $\pmb{\alpha}$ always attributes highest similarity scores to dysarthric speakers rather than healthy ones, when the target speaker is dysarthric, and vice versa for healthy target speakers. This suggests that this approach can realistically estimate speaker closeness.
Fig. \ref{fig:alpha_M13} shows the recovered weights for a dysarthric target speaker. The mass is spread along dysarthric speakers while the $\alpha_{j}$ values are close to zero for healthy speakers.

\subsection{The $\pmb{\alpha}$ weight and the dysarthria detection}
As we showed in Fig. \ref{fig:alpha_M13}, MSDA-WJDOT associates speakers with similar voice characteristics. As additional analysis, we investigated the possibility of leveraging the $\pmb{\alpha}$ weights of the command classifier to detect dysarthria. Specifically, we attempt to classify a speaker as healthy or dysarthric based on his/her similarity with the other subjects.

Let define $I_{c}$ as the set indexing the control speakers and $I_{d}$ as the set of indices related to dysarthric speakers. We then define the Healthy Score (HS) and the Dysarthric Score (DS) as follow 
\begin{equation}
    HS = \sum _{j\in I_{c}} \alpha _{j}, \quad DS = \sum _{j\in I_{d}} \alpha _{j}.
\end{equation}

\begin{figure}[t]
\centering\includegraphics[width=.9\linewidth]{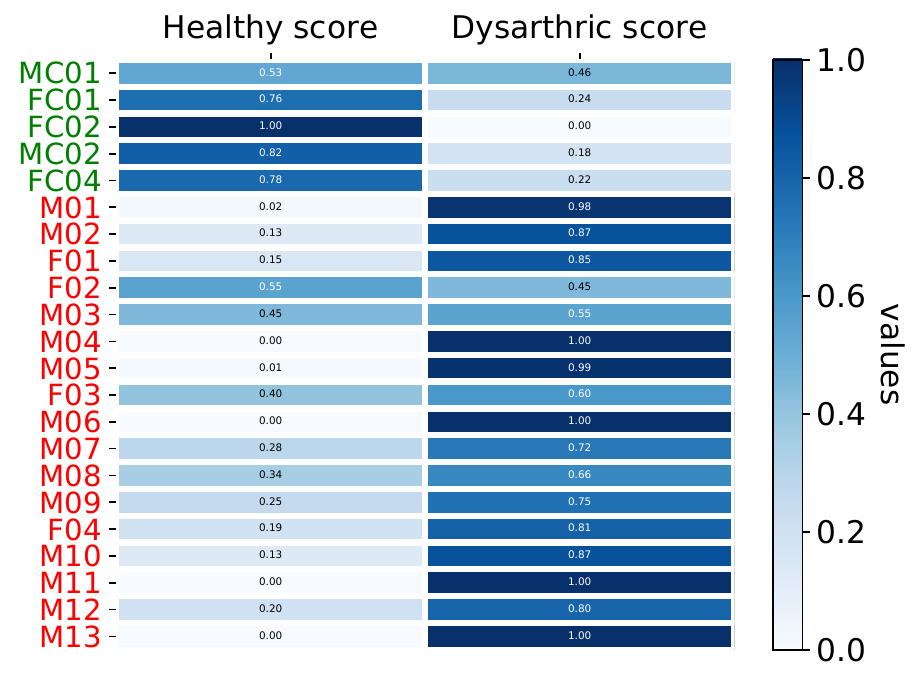}
\caption{HS and DS computed for healthy (in green) and dysarthric (in red) speakers.}
\label{fig:alpha_all_scores}
\end{figure}

We can use these scores to perform dysarthria detection by stating that
$$ A\; speaker\; is\; affected\; by\; dysarthria\; if\; DS > HS.$$
Fig.  \ref{fig:alpha_all_scores} reports the computed scores for all dysarthric speakers and for 5 control speakers. As we can see the controls subjects are always classified as healthy while for the patients, except for F02, we have $DS>HS$. This results in a final accuracy of 95$\%$. 

\section{Conclusions}\label{sec:discussion}
In this work, we addressed usupervised domain and speaker adaptation in the challenging context of pathological speech by employing MSDA-WJDOT \cite{turrisi2020multi}. We have shown the effectiveness of our proposed method on dysarthria detection and spoken command recognition, by comparing it with with well-established MSDA approches and a Baseline model, in which a global classifier is learned on the source datasets and directly tested on the target data, without adaptation. 
Our method provides the best performance on both applications. Interestingly, MSDA-WJDOT also provides source-target similarity coefficients $\pmb{\alpha}$ that, in speaker adaptation, result in a  measure of speaker relatedness.
%Our method outperforms all  the competitors in dysarthria detection task from noisy signals. }In unsupervised adaptation for spoken command recognition our MSDA-WJDOT algorithm  provided the best performance with Command Error Rate reduction of 16$\%$ over the Baseline. Interestingly, our approach is based on the estimation of source-target similarity coefficients $\pmb{\alpha}$ that are a measure of speaker relatedness.
From this, we derived the Healthy (HI) and Dysarthric (DI) Index of a target speaker and diagnosed the dysarthria, achieving an accuracy of 95$\%$.

Future directions could delve into the dysarthria assessment by individuating intervals of values in which the DI corresponds to dysarthria severity levels (e.g., mild, moderate, severe).% that corresponds to dysarthria severity levels, %such that DI$\in[0, v_{1}]$=healthy, DI$\in(v_{1}, v_{2}]$=mild dysarthria, $\dots$, DI$\in(v_{N}, 1]$=severe dysarthria. 
This may bring to very efficient ASR systems that simultaneously improve their performance via SA, and compute the DI warning the subject when the index is close to the right endpoint of its interval. Such a device could predict the disease degeneration and allow the patient to act in time in order to prevent it.

%\section{Acknowledgements}
%This work has been partially funded through the grant from SAP SE and 5x1000, assigned to the University of Ferrara (2017).

\bibliographystyle{IEEEtran}

\bibliography{mybib}

% \begin{thebibliography}{9}
% \bibitem[1]{Davis80-COP}
%   S.\ B.\ Davis and P.\ Mermelstein,
%   ``Comparison of parametric representation for monosyllabic word recognition in continuously spoken sentences,''
%   \textit{IEEE Transactions on Acoustics, Speech and Signal Processing}, vol.~28, no.~4, pp.~357--366, 1980.
% \bibitem[2]{Rabiner89-ATO}
%   L.\ R.\ Rabiner,
%   ``A tutorial on hidden Markov models and selected applications in speech recognition,''
%   \textit{Proceedings of the IEEE}, vol.~77, no.~2, pp.~257-286, 1989.
% \bibitem[3]{Hastie09-TEO}
%   T.\ Hastie, R.\ Tibshirani, and J.\ Friedman,
%   \textit{The Elements of Statistical Learning -- Data Mining, Inference, and Prediction}.
%   New York: Springer, 2009.
% \bibitem[4]{YourName17-XXX}
%   F.\ Lastname1, F.\ Lastname2, and F.\ Lastname3,
%   ``Title of your INTERSPEECH 2021 publication,''
%   in \textit{Interspeech 2021 -- 20\textsuperscript{th} Annual Conference of the International Speech Communication Association, September 15-19, Graz, Austria, Proceedings, Proceedings}, 2020, pp.~100--104.
% \end{thebibliography}

\end{document}